\documentclass[twocolumn,aps,prd,reprint,superscriptaddress,nofootinbib,letterpaper]{revtex4}

\usepackage{amsmath}
\usepackage{amsfonts}
\usepackage{amssymb}
\usepackage{bm} 
\usepackage{graphicx}
\usepackage{epsfig}
\usepackage{color}
\usepackage{multirow}
\usepackage{todonotes}
\usepackage{hyperref}

\def\bea{\begin{eqnarray}}
\def\eea{\end{eqnarray}}
\def\bef{\begin{flalign}}
\def\eef{\end{flalign}}
\def\nn{\nonumber}
\def\d{\mathrm{d}}

\def\vx{\vec{x}}

\def\({\left(}
\def\){\right)}
\def\[{\left[}
\def\]{\right]}
\def\<{\left\langle}
\def\>{\right\rangle}

\newcommand{\icol}[1]{
\left[\begin{smallmatrix}#1\end{smallmatrix}\right]%
}

\begin{document}


\title{\center{Thermalization and localization in a discretized quantum field theory}}

\author{Spasen Chaykov}
\email{spasen\_chaykov@student.uml.edu}
\affiliation{Department of Physics and Applied Physics, University of Massachusetts, Lowell, MA 01854, USA}

\author{Brenden Bowen}
\email{brenden\_bowen@student.uml.edu}
\affiliation{Department of Physics and Applied Physics, University of Massachusetts, Lowell, MA 01854, USA}

\author{Nishant Agarwal}
\email{nishant\_agarwal@uml.edu}
\affiliation{Department of Physics and Applied Physics, University of Massachusetts, Lowell, MA 01854, USA}

\date{\today}

\begin{abstract}
Localization marks the breakdown of thermalization in subregions of quantum many-body systems in the presence of sufficiently large disorder. In this paper, we use numerical techniques to study thermalization and localization in a many-body system of coupled quantum harmonic oscillators obtained by discretizing a scalar quantum field theory in Minkowski spacetime. We consider a Gaussian initial state, constructed through a global mass quench, with a quadratic Hamiltonian, and solve for the system's exact dynamics without and with disorder in one and two spatial dimensions. We find that finite-size systems localize for sufficiently large disorder in both cases, such that the entanglement entropy of subregions retains its initial area-law behavior, and the system no longer develops long-range correlations. To probe the thermalization-to-localization transition further, we define a frequency gap ratio that measures adjacent gaps in the phase space eigenvalues of the Hamiltonian and study how it varies with disorder strength and system size. We find signatures of a chaotic regime at intermediate disorder in two spatial dimensions and argue that it is a finite-size effect, such that the system would localize for arbitrarily small disorder in the continuum in both one and two spatial dimensions, consistent with Anderson localization. Lastly, we use the frequency gap ratio to argue that in three spatial dimensions, on the other hand, the system would only localize for disorder strengths above a critical value in the continuum, again consistent with Anderson localization.
\end{abstract}

\maketitle


\section{Introduction}
\label{sec:intro}

The eigenstate thermalization hypothesis \cite{Deutsch:1991,Srednicki:1994mfb,Rigol:2008,DAlessio:2015qtq} suggests that local observables in isolated quantum many-body systems evolve in time and equilibrate to their canonical ensemble expectation values, bringing, in particular, subregions to thermal equilibrium. Introducing sufficiently large disorder in the Hamiltonian can, however, localize excitations \cite{Anderson:1958vr}, thus preventing the system from thermalizing. In contrast to thermalizing systems, those in the localized phase do not develop long-range correlations, and subregions retain memory of their initial entanglement structure. Localization is also characterized by unique spectral statistics in measures such as the spectral gap ratio and spectral form factor \cite{Oganesyan:2007,Pal:2010,Prakash:2020psj,Suntajs:2021qqr,Suntajs:2022ldk}, that have interesting connections to random matrix theory as well \cite{Haake:2010,Atas:2013}.

Whereas finite-size systems are expected to localize for disorder strengths above some critical value, whether and how systems localize in the thermodynamic limit depends on the details of the system Hamiltonian. In the absence of interactions (Anderson localization), one- and two-dimensional systems localize for infinitesimally small disorder in the thermodynamic limit, while three-dimensional systems localize only for disorder strengths above a critical value \cite{Anderson:1958vr, Abrahams:1979, ATAF:1980, Lee:1985}. In the presence of interactions (many-body localization), on the other hand, recent work suggests that systems may not localize in the thermodynamic limit since the transition shifts to larger disorder strengths as one increases the system size \cite{Suntajs:2019lmb,Suntajs:2020,Morningstar:2021pcy,Sels:2021rjm,Sierant:2021jay} and impurities continue to relax as time evolves \cite{Kiefer:2021,Sels:2021azw}.

In this paper, we are interested in understanding thermalization and localization in a discretized quantum field theory (QFT), specifically a Klein-Gordon field theory in Minkowski spacetime in one, two, and three spatial dimensions, and whether a quantum field can be localized in the continuum limit. We start by discretizing the QFT on a spatial lattice, which reduces it to a many-body system of quantum harmonic oscillators with nearest-neighbor interactions of a particular form. The resulting Hamiltonian is quadratic, and we add to it a disorder term of the usual on-site number operator form that is also quadratic. The Hamiltonian in both the absence and presence of disorder can then be diagonalized exactly in phase space. Further, choosing a Gaussian initial state that we construct through a global mass quench allows us to solve for the system's {\it exact} dynamics (using numerical techniques). Another advantage of working in phase space, as this many-body system permits, is that the dimension of the phase space Hamiltonian matrix scales polynomially with the number of oscillators, even though the dimension of the Hilbert space scales exponentially.

We solve for the system's dynamics in one and two spatial dimensions and find, as expected, that subregions thermalize in the absence of disorder, with the von Neumann entanglement entropy transitioning from the corresponding area-law \cite{Bombelli:1986,Srednicki:1993im} to volume-law behavior \cite{Calabrese:2005evo,Cotler:2016acd}. In the presence of sufficiently large disorder, on the other hand, we find that the entanglement entropy of subregions freezes to its initial area-law behavior. Additionally, the two-point correlation function decays rapidly with distance, suggesting that the system has localized. To probe the thermalization-to-localization transition further, we define a frequency gap ratio that measures adjacent gaps in the phase space eigenvalues of the Hamiltonian, analogous to the spectral gap ratio defined in \cite{Oganesyan:2007} that measures adjacent gaps in the energy eigenvalues of the Hamiltonian. The frequency gap ratio allows us to work around the infinite energy spectrum of our many-body system, and we expect it to be a reliable measure of localization since the system's dynamics are described fully by its frequency spectrum. We use it to argue that the system would localize for arbitrarily small disorder in the continuum in both one and two spatial dimensions, even though the two-dimensional system exhibits signatures of chaos for finite system sizes. Lastly, we use the frequency gap ratio to show that in three spatial dimensions, the system would only localize for disorder strengths above a critical value in the continuum. Our results suggest that it is possible to localize the simple QFT we consider using infinitesimal disorder in one and two spatial dimensions and finite disorder in three, consistent with Anderson localization.

The paper is organized as follows. We first describe the discretized Hamiltonian both without and with disorder in section\ \ref{sec:hamdisc}. We next discuss the covariance matrix approach to calculating entanglement entropy in section\ \ref{sec:ee}, relegating details on the diagonalization procedure and time evolution to two appendices. In section\ \ref{sec:results}, we present our results on thermalization and localization, highlighting the differences between one and two spatial dimensions. We then define the frequency gap ratio and compare its behavior in one, two, and three spatial dimensions in section\ \ref{sec:fgr} and end with a discussion in section\ \ref{sec:disc}.


\section{Discretized Hamiltonian}
\label{sec:hamdisc}

Consider a free scalar field $\hat{\phi}(\vx,t)$ in $(d+1)$-dimensional Minkowski spacetime with the Hamiltonian
\bea
    \hat{H} & = & \frac{1}{2} \int \d^dx \left[ \hat{\pi}^2 + \big(\vec{\nabla} \hat{\phi}\big)^2 + m^2 \hat{\phi}^2 \right] ,
\label{eq:ham}
\eea
where $\hat{\pi}(\vx,t)$ is the momentum conjugate to $\hat{\phi}(\vx,t)$ and $m$ is the mass of the field. We first transition to a finite-size and discretized theory where the field turns into a collection of $N$ harmonic oscillators arranged on a $d$-dimensional lattice, with $N^{1/d}$ oscillators on each side. This amounts to the substitutions 
\bea
    \hat{\phi}(\vx,t) & \rightarrow & \epsilon^{-(d-1)/2} \, \hat{\phi}_i(t) \, , \\
    \hat{\pi}(\vx,t) & \rightarrow & \epsilon^{-(d+1)/2} \, \hat{\pi}_i(t) \, , \\
    \int \d^dx & \rightarrow & \epsilon^d \sum_{i=1}^N \, ,
\label{eq:1dsubs}
\eea
where $\epsilon$ is the lattice spacing, the index $i$ (and $j$ below) runs over all lattice sites, and we have defined dimensionless operators $\hat{\phi}_i$ and $\hat{\pi}_i$ by including appropriate factors of $\epsilon$ in their definitions. Making these substitutions in eq.\ (\ref{eq:ham}) gives us the discretized Hamiltonian
\bea
    \hat{H} & = & \frac{1}{2\epsilon} \sum_{i=1}^{N} \big( \hat{\pi}_i^2 + m^2 \epsilon^2 \hat{\phi}_i^2 \big) + \frac{1}{2\epsilon} \sum_{\langle i j \rangle_b} \big( \hat{\phi}_{i} - \hat{\phi}_{j} \big)^2 \, , \quad
\label{eq:hamdisc}
\eea
where angular brackets on the second sum indicate that it runs only over nearest-neighbor connections on the lattice and the subscript $b$ denotes the boundary conditions imposed on the system. We impose Dirichlet boundary conditions such that oscillators on the edges are connected to a `wall', resulting in additional $\hat{\phi}_i^2/(2\epsilon)$ terms for these oscillators.

\begin{figure*}[!t]
\begin{center}
    \begin{tikzpicture}
        \draw[color=white,opacity=0] (0,-2) rectangle (4,2);
        \foreach \x in {0,...,2}{
            \draw[gray] (\x,0) -- ++(1,0);
        };
        \draw[gray] (3,0) -- ++(0.5,0);
        \foreach \x in {1,...,3}{
            \shade[ball color=black] (\x,0) circle (.15);
        };
        \draw[thick, dashed, blue] (0,0.5) -- (2.5,0.5) -- (2.5,-0.5) -- (0,-0.5);
        \node[below,blue] at (1.25,-0.5){$ L $};
        \foreach \y in {1,...,5}{
            \draw[black,ultra thick] (0,\y/2.5-1) -- (-0.5,\y/2.5-1.4);
        };
        \draw[ultra thick] (-0.5,-1) rectangle (0,1);
    \end{tikzpicture}
    \hspace{80pt}
    \begin{tikzpicture}
        \draw[step=1,color=gray] (0.5,0.5) grid (4.5,4.5);
        \foreach \x in {1,...,4}
            \foreach \y in {1,...,4}{
            \shade[ball color=black] (\x,\y) circle (.15);
        };
        \draw[thick, dashed, blue] (1.5,1.5) -- (3.5,1.5) -- (3.5,3.5) -- (1.5,3.5) -- cycle;
        \node[below,blue] at (2.5,1.5){$ L $};
    \end{tikzpicture}
    \caption{Lattice setup in one and two spatial dimensions. (Left) A chain of oscillators with nearest-neighbor interactions and strip subregions chosen from the left edge of the chain. The `wall' at the left edge indicates Dirichlet boundary conditions, and the number of oscillators within a subregion of size $L$ is $L/\epsilon - 1/2$. (Right) A two-dimensional lattice of oscillators again with nearest-neighbor interactions and square subregions chosen in the middle of the full lattice. The number of oscillators on the side of a square of size $L$ is now $L/\epsilon$.}
\label{fig:geom}
\end{center}
\end{figure*}
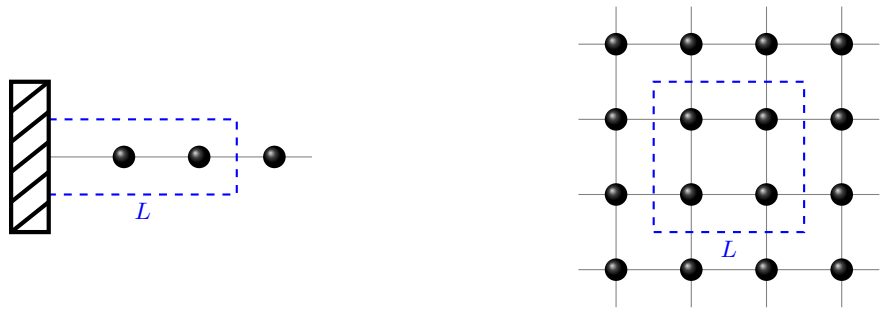

\begin{figure*}[!t]
\begin{center}
    \includegraphics[scale=0.55]{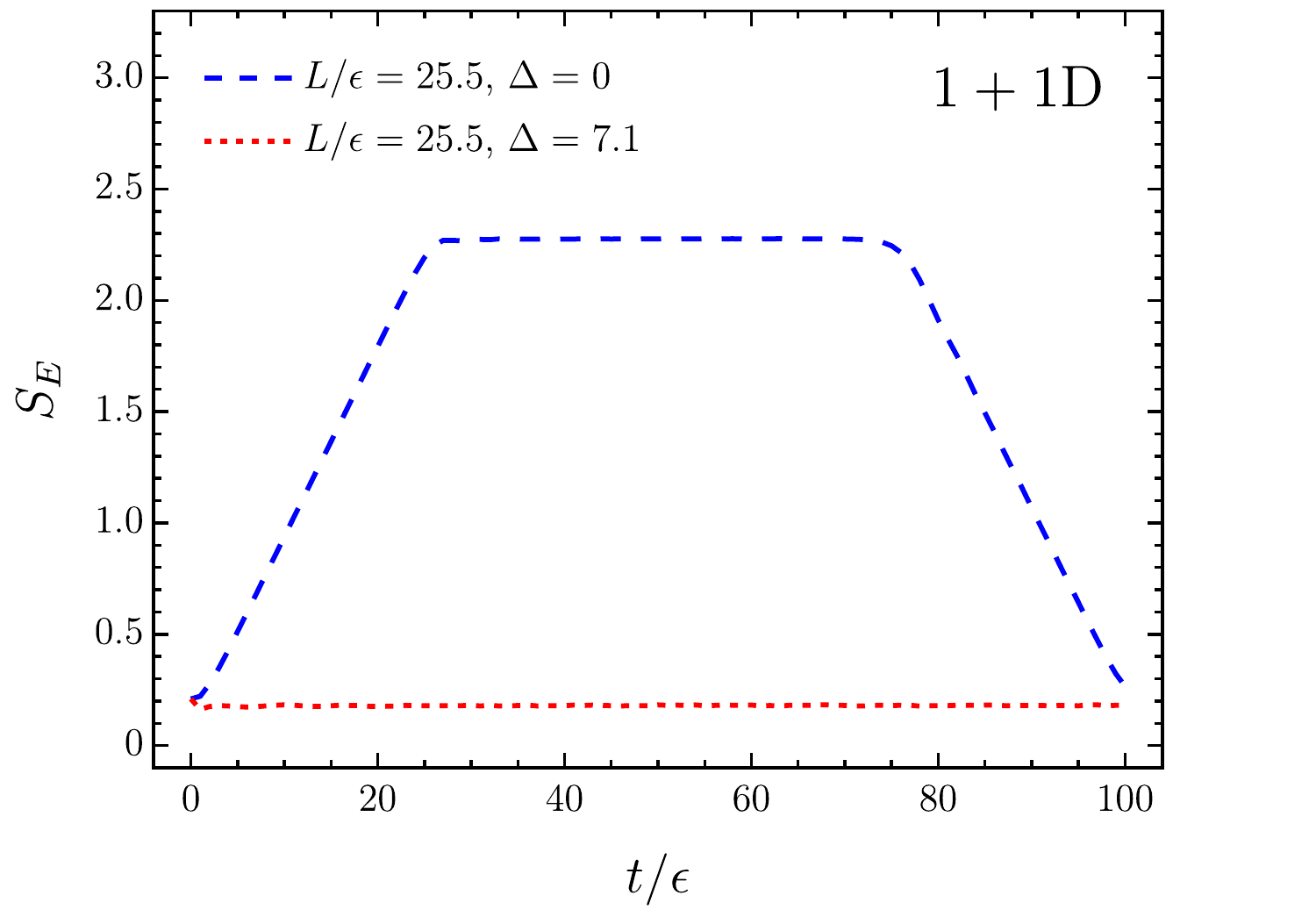}
    \includegraphics[scale=0.55]{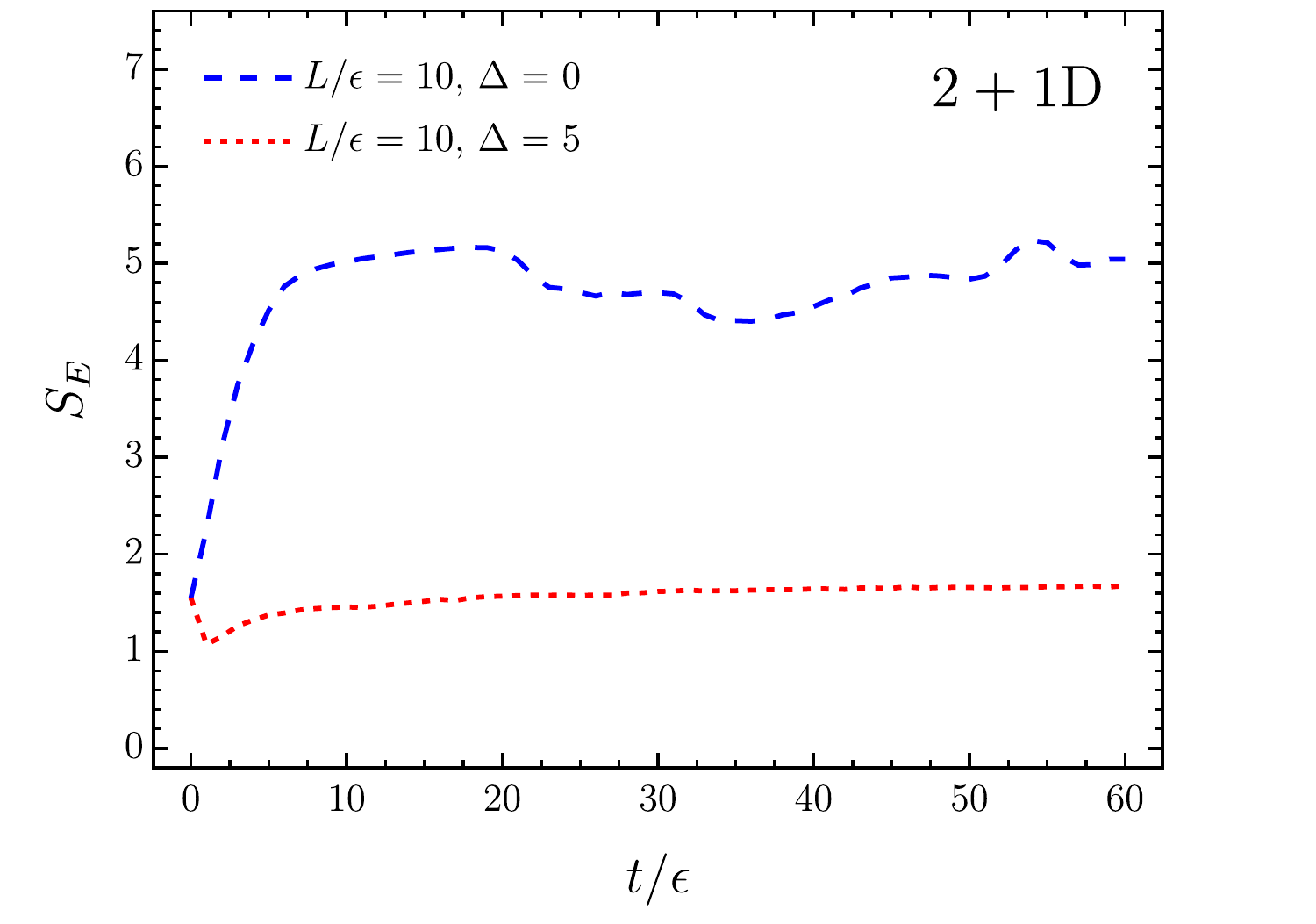}
    \caption{Entanglement entropy as a function of time for a given subregion size, both without and with disorder, following a global mass quench. (Left) For a chain of oscillators with $N = 100$, $m = 0.3/\epsilon$, $\Delta = 10/\omega_0$, $\omega_0 = \sqrt{2}$, and $L/\epsilon = 25.5$. (Right) For a two-dimensional lattice of oscillators with $N = 50^2$, $m = 0.6/\epsilon$, $\Delta = 10/\omega_0$, $\omega_0 = 2$, and $L/\epsilon = 10$.}
\label{fig:2}
\end{center}
\end{figure*}

The Hamiltonian in eq.\ (\ref{eq:hamdisc}) can be written in the form
\bea
    \hat{H} & = & \frac{1}{2\epsilon} \hat{\boldsymbol{\chi}}^{T} \boldsymbol{V} \hat{\boldsymbol{\chi}} \, ,
\label{eq:Hamapp} 
\eea
where $\hat{\boldsymbol{\chi}} = \icol{\hat{\boldsymbol{\phi}} \\ \hat{\boldsymbol{\pi}}}$ is the phase space vector in the physical basis, with
\bea
\label{eq:BoldPhiPi}
    \hat{\boldsymbol{\phi}} \, = \, \[
    \begin{array}{c}
		\hat{\phi}_1 \\
		\vdots \\
		\hat{\phi}_N
	\end{array} \] \ \ {\rm and} \quad
    \hat{\boldsymbol{\pi}} \, = \, \[
    \begin{array}{c}
		\hat{\pi}_1 \\
		\vdots \\
		\hat{\pi}_N
	\end{array} \] ,
\eea
$\boldsymbol{V}$ is a $2N \times 2N$ matrix whose diagonal terms constitute the free part of the Hamiltonian and off-diagonal terms the couplings between different oscillators, and the superscript $T$ indicates a transpose. For eq.\ (\ref{eq:hamdisc}), the Hamiltonian matrix $\boldsymbol{V}$ is given by
\bea
    \boldsymbol{V} & = &  \[
    \begin{array}{cc}
        \boldsymbol{K} & 0 \\
        0 & \boldsymbol{I}_N
    \end{array} \] , 
\label{eq:V1d}
\eea
where $\boldsymbol{I}_N$ is the $N$-dimensional identity matrix and $\boldsymbol{K}$ is the Laplacian matrix given in turn by
\bea
    K_{ij} \ = \ 
    \begin{cases}
        \deg(v_i)+m^2\epsilon^2 & \textnormal{if } i = j \\
        -1 & \textnormal{if } v_i \textnormal{ is adjacent to } v_j \\
        \phantom{-}0 & \textnormal{otherwise} \, ,
    \end{cases}\label{eq:Kmat}
\eea
where $v_i$ indicates vertices on the lattice and $\deg(v_i)$ is the number of connections to a given vertex. Note that $\deg(v_i)$ is the same for all oscillators in a given number of spatial dimensions for our choice of boundary conditions. We show the lattice setup in one ($d = 1$) and two ($d = 2$) spatial dimensions in fig.\ \ref{fig:geom} for clarity, additionally showing there the subregions whose entanglement entropy will be calculated in the next section. In the first case, the oscillators are arranged on a chain and have at most two nearest neighbors so that $\deg(v_i)$ is two, and in the second case, they are arranged on a square lattice and have at most four nearest neighbors, so that $\deg(v_i)$ is four.\footnote{We note that our lattice construction differs from the radial lattice approach used in \cite{Srednicki:1993im}. The two methods lead to a qualitatively different time-dependence in the entanglement entropy that we comment on later in the context of fig.\ \ref{fig:2}.}

We next add a local disorder Hamiltonian of the usual on-site number operator form to eq.\ (\ref{eq:hamdisc}),
\bea
    \hat{H}_{\text{disorder}} & = & \frac{1}{\epsilon} \sum_{i=1}^{N} h_i \hat{a}^{\dagger}_i \hat{a}_i \, ,
\label{eq:hamdis}
\eea
where $\hat{a}_i$ ($\hat{a}_i^{\dagger}$) are annihilation (creation) operators at each lattice site and $h_i$ are random numbers. The $\hat{a}_i$ here are related to $\hat{\phi}_i$ and $\hat{\pi}_i$ as $\hat{a}_{i} = \sqrt{\omega_0/2} \big(  \hat{\phi}_{i} + i \hat{\pi}_{i}/\omega_0 \big)$, where $\omega_0 = \sqrt{\deg(v_i)+m^2\epsilon^2}$ is the dimensionless frequency of the non-interacting part of the lattice, so that $\omega_0 = \sqrt{2+m^2\epsilon^2}$ and $\omega_0 = \sqrt{4+m^2 \epsilon^2}$ in one and two spatial dimensions, respectively. The disorder term, therefore, introduces additional terms involving both the field and conjugate momentum operators, resulting in a modification of the diagonal components of $\boldsymbol{V}$ in eq. (\ref{eq:V1d}). For spin systems, the $h_i$ are usually chosen from a uniform distribution in $[-\Delta,\Delta]$, though systems with correlated disorder also exhibit interesting localization properties \cite{De:1998,Shima:2004,singh:2021,Shi:2022}. In the many-body system we consider, negative values of $h_i$ can lead to Hamiltonian instabilities and we, therefore, draw the $h_i$ from a uniform distribution in $[0,\Delta]$ instead. Since the exact set of $h_i$ varies for different realizations of the Hamiltonian, we refer to $\Delta$ as the strength of the disorder term. We also note that the disorder term in eq.\ (\ref{eq:hamdis}) differs from the one considered in \cite{Abdul-Rahman:2021bmd}, which is closer to a mass disorder term.


\section{Entanglement entropy}
\label{sec:ee}

We next introduce dynamics by choosing the initial state of our many-body system to be a non-eigenstate of the Hamiltonian written in the previous section. Specifically, we choose the initial state to be the ground state of a massive Hamiltonian, i.e., that in eq.\ (\ref{eq:hamdisc}) with $m \ne 0$, and then evolve with either the mass{\it less} Hamiltonian, i.e., that in eq.\ (\ref{eq:hamdisc}) with $m = 0$, or by the massless Hamiltonian plus the disorder Hamiltonian in eq.\ (\ref{eq:hamdis}). This {\it global mass quench} has the advantage that it generates a Gaussian initial state. Since the Hamiltonian is also quadratic, the dynamics are fully determined by the two-point correlations of the system or, equivalently, the covariance matrix
\bea
    \Gamma_{AB}(t) & = & \frac{1}{2} \big\langle \{\hat{\chi}_A(t), \hat{\chi}_B(t)\} \big\rangle-\big\langle \hat{\chi}_A(t)\big\rangle\big\langle \hat{\chi}_B(t) \big\rangle \, ,\nn\\
\label{eq:gammadef}
\eea
where the indices $A$ and $B$ run from $1$ to $2N$ and $\{\cdot,\cdot\}$ is the anti-commutator.

\begin{figure*}[!t]
\begin{center}
	\includegraphics[scale=0.55]{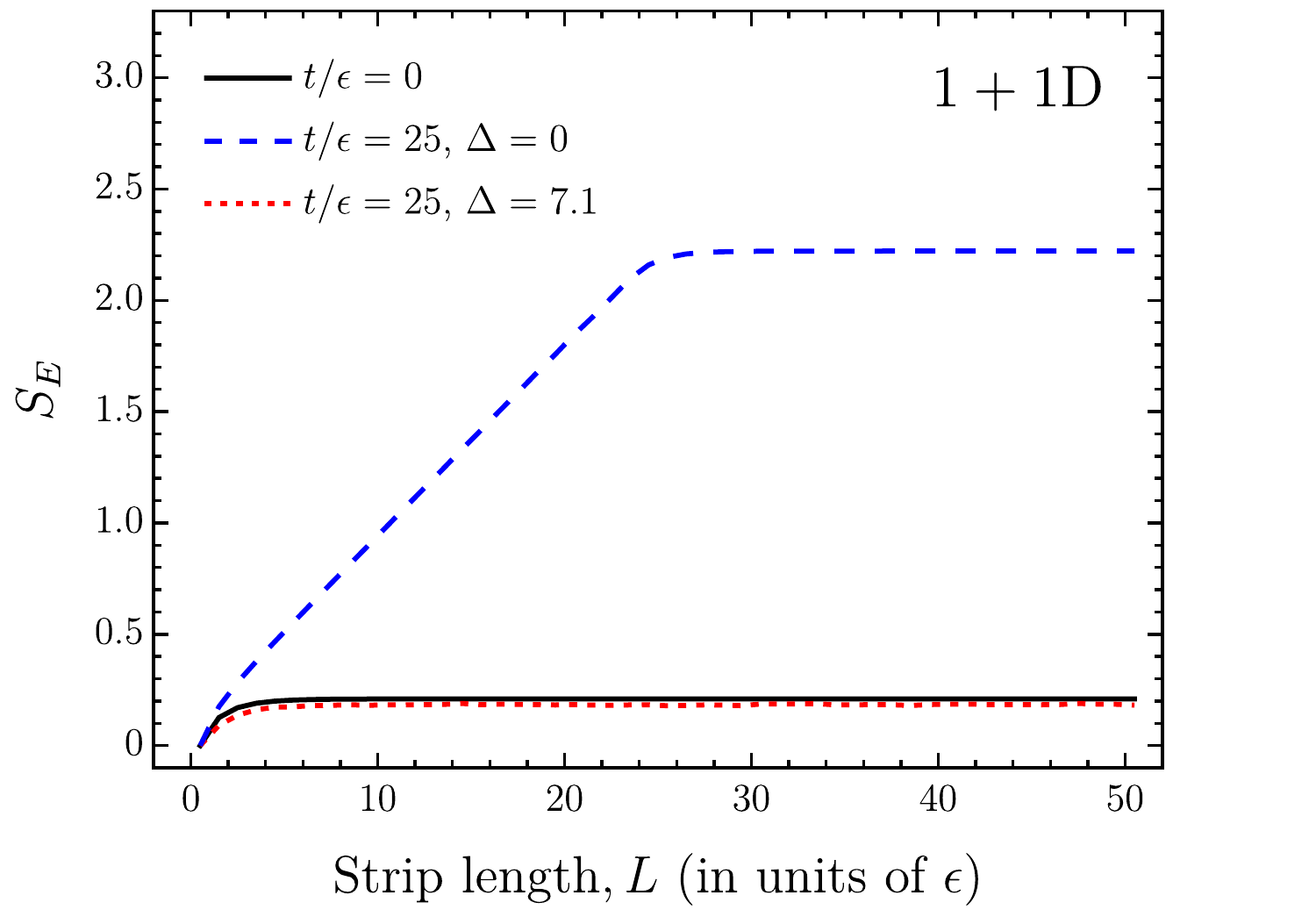}
	\includegraphics[scale=0.55]{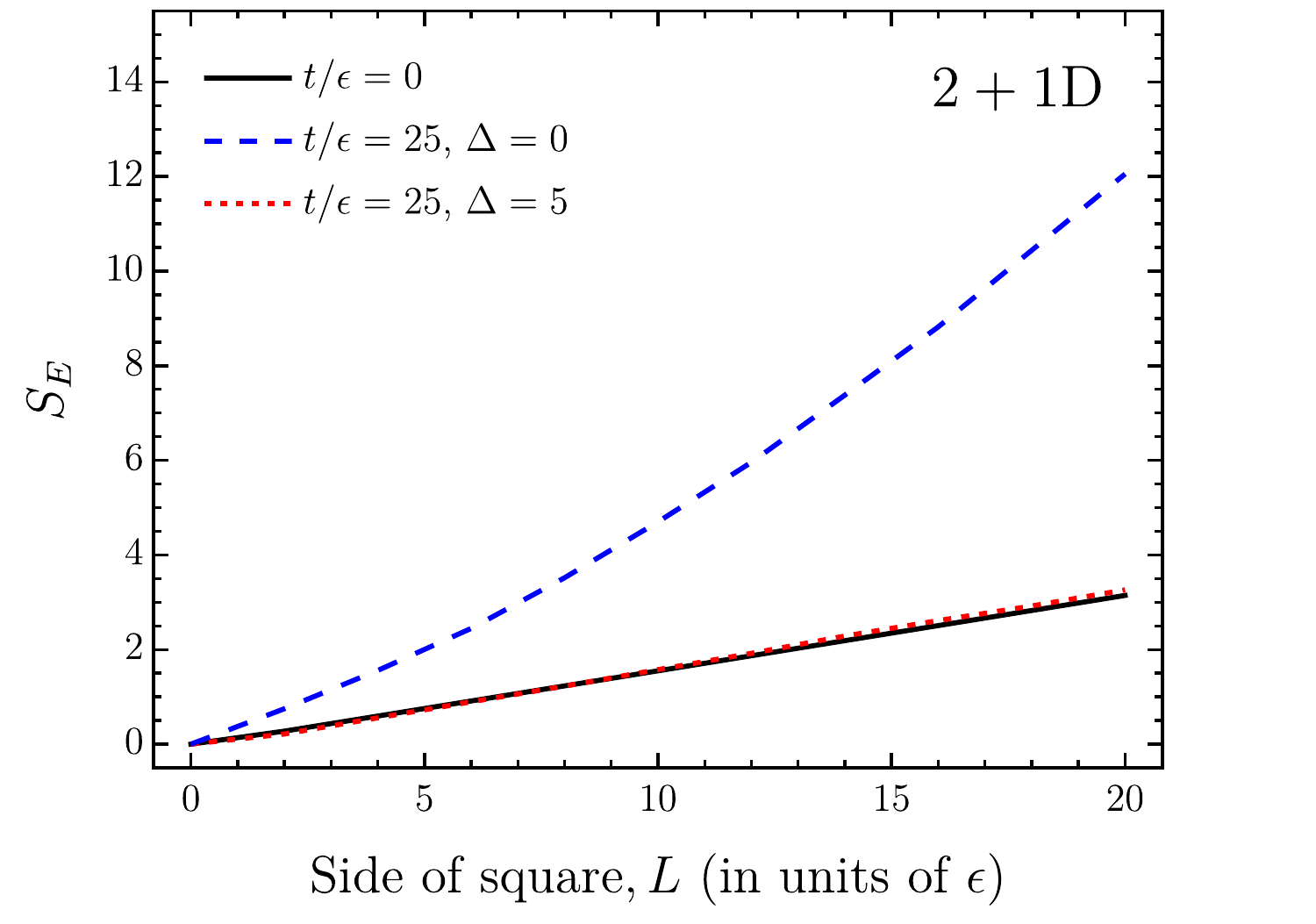}
	\caption{Entanglement entropy as a function of subregion size at the initial time and $t/\epsilon = 25$, both without and with disorder, following a global mass quench. (Left) For a chain of oscillators with $N = 100$, $m = 0.3/\epsilon$, $\Delta = 10/\omega_0$, and $\omega_0 = \sqrt{2}$. (Right) For a two-dimensional lattice of oscillators with $N = 50^2$, $m = 0.6/\epsilon$, $\Delta = 10/\omega_0$, and $\omega_0 = 2$.}
\label{fig:1}
\end{center}
\end{figure*}

In order to construct the ground state of the massive Hamiltonian and also to determine the system's dynamics, we first need to diagonalize the Hamiltonian matrix $\boldsymbol{V}$. Since $\boldsymbol{V}$ is symmetric and positive-definite, it can be brought to a diagonal Williamson form $\boldsymbol{W}$ by means of a symplectic transformation matrix $\boldsymbol{M}$, such that $\boldsymbol{V} = \boldsymbol{M}^{T} \boldsymbol{W} \boldsymbol{M}$. We describe this phase space diagonalization and the procedure to find $\boldsymbol{M}$ for a general quadratic Hamiltonian in appendix\ \ref{app:diag}. Using this method, the Hamiltonian in eq.\ (\ref{eq:Hamapp}) can be written in the decoupled form
\bea
    \hat{H} & = & \frac{1}{2\epsilon} \hat{\boldsymbol{\chi}}_{D}^{T} \boldsymbol{W} \hat{\boldsymbol{\chi}}_D  \\
    & = & \frac{1}{2\epsilon} \sum_{i=1}^N \omega_i \big( \hat{\pi}_{D,i}^2 + \hat{\phi}_{D,i}^2 \big) \, ,
\label{eq:Hamdiag}
\eea
where $\hat{\boldsymbol{\chi}}_D = \boldsymbol{M} \hat{\boldsymbol{\chi}}$ is the phase space vector in the decoupled basis and $\omega_i$ are the symplectic eigenvalues\footnote{As discussed in appendix\ \ref{app:diag}, a $2N \times 2N$ symmetric positive-definite matrix can be written in the diagonal Williamson form, where the diagonal entries are two copies of the $N$ symplectic eigenvalues.} of $\boldsymbol{V}$. Once we have determined the decoupled basis for the massive Hamiltonian, we can construct the ground state of the system by taking a tensor product of the ground states of individual decoupled modes. We can then obtain the two-point correlations in this state to construct the initial covariance matrix in the decoupled basis and finally transform back to the physical basis to obtain $\boldsymbol{\Gamma}(0)$.

Now that we have the initial state in the physical basis, we want to time evolve it with either the massless Hamiltonian or the massless Hamiltonian plus the disorder Hamiltonian, as mentioned earlier. Since time evolution is trivial in the decoupled basis, we again start by finding the decoupled basis of the new Hamiltonian using the method described in appendix\ \ref{app:diag}. We then transform the prepared initial state $\boldsymbol{\Gamma}(0)$ to the decoupled basis of the new Hamiltonian, $\boldsymbol{\Gamma}_D(0)$, time evolve to find $\boldsymbol{\Gamma}_D(t)$, and finally transform back to the physical basis to obtain $\boldsymbol{\Gamma}(t)$. We describe the time evolution of the covariance matrix starting in a general Gaussian initial state in more detail in appendix \ref{app:td}. We also note that while it is possible to diagonalize the discretized Hamiltonian in eq.\ (\ref{eq:hamdisc}) simply through an orthogonal transformation on $\hat{\boldsymbol{\phi}}$, the more general phase space diagonalization described here is imperative once we include the disorder Hamiltonian since it introduces nontrivial terms in the conjugate momentum sector of $\boldsymbol{V}$.

We are now ready to obtain the entanglement entropy for subregions of our many-body system. As also shown in fig.\ \ref{fig:geom}, in the case of one spatial dimension, we choose strip subregions from the left edge of the chain, and in the case of two spatial dimensions, we choose square subregions in the middle of the full lattice. We denote both the strip length and the side of the square with $L$ as this is the relevant length scale in both cases. In either case, let us say that the subregion consists of $n$ oscillators. The entanglement entropy of this subregion can then be obtained by diagonalizing the corresponding $2n \times 2n$ submatrix of $\boldsymbol{\Gamma}(t)$ and using the formula \cite{Eisert.82.277}
\bea
    S_E(t) & = & \sum_{i=1}^{n} \bigg[ \left( \gamma_i + \frac{1}{2} \right) \ln \left( \gamma_i + \frac{1}{2} \right) \nn \\
    & & \hspace{25pt} - \, \left( \gamma_i - \frac{1}{2} \right) \ln \left( \gamma_i - \frac{1}{2} \right) \bigg] \, , \quad
\label{eq:eedef}
\eea
where $\gamma_i = \gamma_i(t)$ are the symplectic eigenvalues of the $2n \times 2n$ submatrix of $\boldsymbol{\Gamma}(t)$. 


\section{Thermalization and localization}
\label{sec:results}

In this section, we use the setup described in the previous two sections to obtain results on the evolution of the entanglement entropy of subregions in one and two spatial dimensions. We first note the various parameter values that we choose. In the one-dimensional case, we consider a total of $N = 100$ oscillators arranged on a chain and discuss the effect of varying $N$ in the next section. For the initial state, which we choose as the ground state of the massive Hamiltonian in eq.\ (\ref{eq:hamdisc}), we set $m = 0.3/\epsilon$. For evolution in the presence of the massless Hamiltonian plus the disorder Hamiltonian, we set the disorder strength to be $\Delta = 10/\omega_0$ with $\omega_0 = \sqrt{2}$, which turns out to be sufficiently large to observe localization and discuss the effect of varying $\Delta$ also in the next section. Lastly, in the presence of disorder, we average our results over $500$ random realizations of the $h_i$. In the two-dimensional case, on the other hand, we consider a total of $N = 50^2$ oscillators arranged on a $50 \times 50$ square lattice, choose $m = 0.6/\epsilon$ for the massive Hamiltonian, set the disorder strength to again be $\Delta = 10/\omega_0$ but now with $\omega_0 = 2$, and average over 50 random realizations of the $h_i$.

In fig.\ \ref{fig:1}, we examine how entanglement entropy depends on subregion size at different times. At the initial time, and in the one-dimensional case, the entanglement entropy obeys the well-known result from \cite{Calabrese:2004eu}, namely $S \sim (1/6) \ln\left( L / \epsilon \right)$, for $Lm \ll 1$ and $S \sim (1/6) \ln [ 1/ (m\epsilon) ]$, for $Lm \gg 1$, where $m^{-1}$ plays the role of the correlation length. In the two-dimensional case, on the other hand, the entanglement entropy follows the expected area law, scaling with the perimeter of the square subregion. As time evolves, we find that it becomes extensive in the absence of disorder, scaling with the size of the strip in the one-dimensional case and the area of the square region in the two-dimensional one. This transition indicates that subregions of the quenched, zero disorder, system thermalize in time. In the presence of local disorder, however, we find that subregions no longer thermalize in the late-time limit, with the scaling of the entanglement entropy remaining strikingly close to that of the initial state, suggesting that the system has localized.

In fig.\ \ref{fig:2}, we show the time evolution of entanglement entropy for a given subregion size. The key feature in the absence of disorder in both the one- and two-dimensional cases is that the entanglement entropy grows linearly after the quench, saturating at $t \approx L$. We note that the one-dimensional system shows a recurrence of the initial state entanglement entropy at $t \approx N \epsilon$, indicating that the apparent thermalization does not persist, though the recurrence time would be pushed to infinity in the thermodynamic limit. Interestingly, the two-dimensional system does not show such a revival, and the entanglement entropy remains roughly constant at late times, with only some noise appearing due to the finite lattice size. The revival {\it would}, however, be present even in the two-dimensional case if we instead discretized the system along a radial direction after decomposing the field using a complete basis of angular functions. This suggests a fundamental difference between the two discretizations, at least for finite-size systems, and how they approach the continuum. As fig.\ \ref{fig:2} shows, in the presence of disorder, on the other hand, the entanglement entropy is roughly constant and saturates to a value close to the initial one.

\begin{figure*}[!t]
\begin{center}
    \includegraphics[scale=0.55]{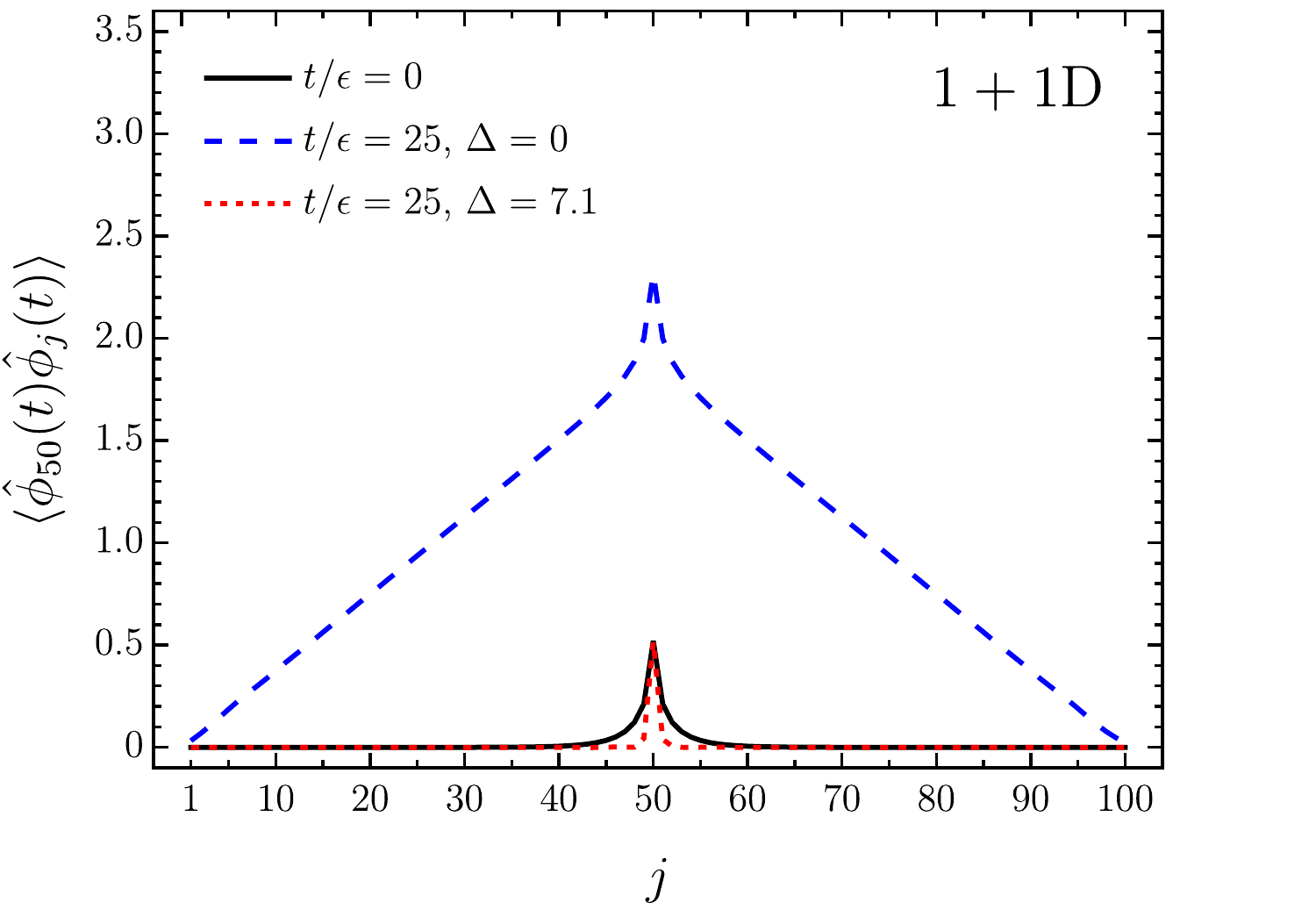}
    \includegraphics[scale=0.55]{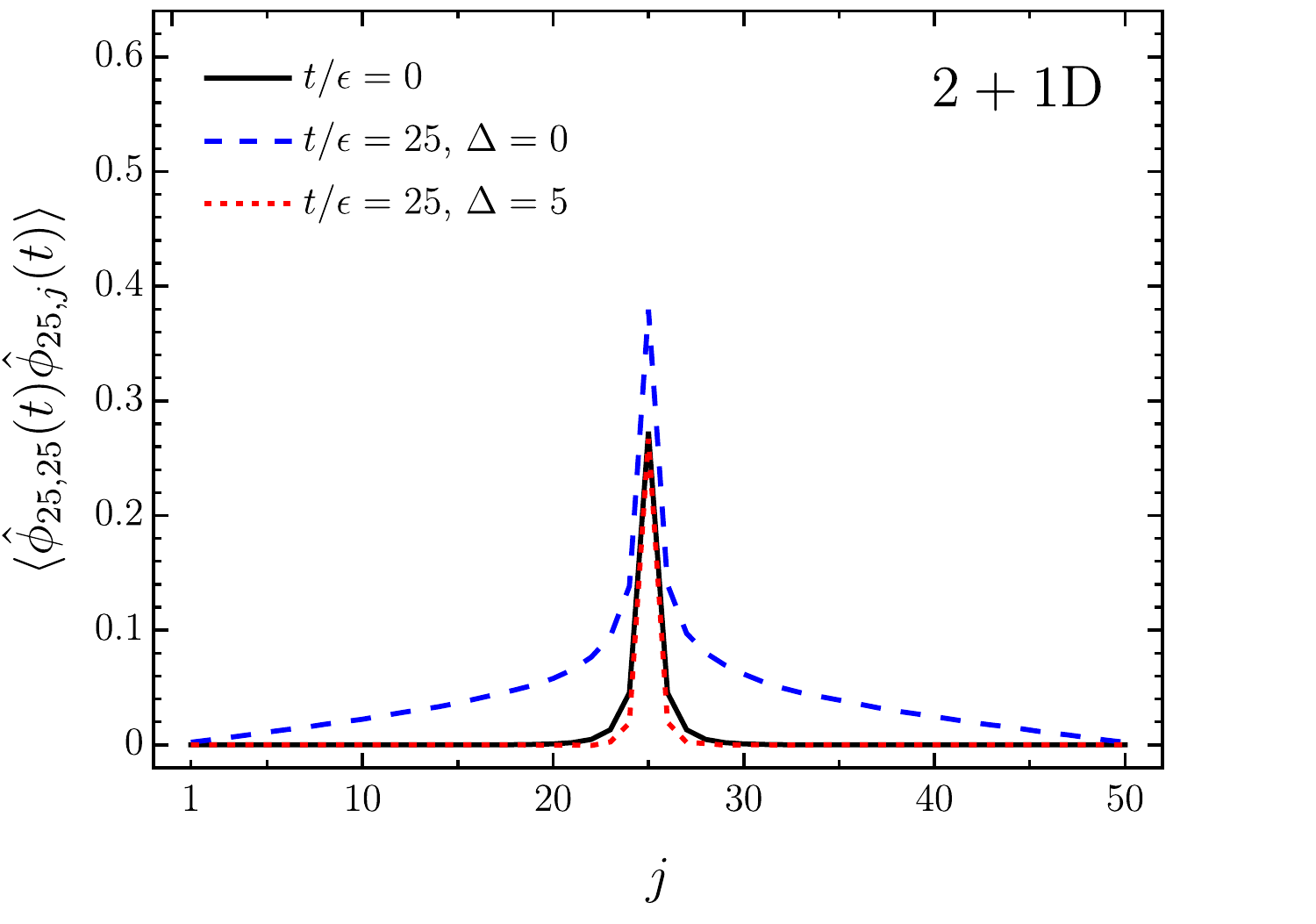}
    \caption{Correlation function as a function of spatial separation at the initial time and at $t/\epsilon = 25$, both without and with disorder, following a global mass quench. (Left) For a chain of oscillators with one oscillator in the middle of the chain and $j$ indicating the position of the other and with $N = 100$, $m = 0.3/\epsilon$, $\Delta = 10/\omega_0$, and $\omega_0 = \sqrt{2}$. (Right) For a two-dimensional lattice of oscillators with one oscillator fixed in the middle of the lattice and $j$ indicating the position of another in the same row and with $N = 50^2$, $m = 0.6/\epsilon$, $\Delta = 10/\omega_0$, and $\omega_0 = 2$.}
\label{fig:3}
\end{center}
\end{figure*}

Lastly, in fig. \ref{fig:3}, we show the behavior of the correlation function $\langle \hat{\phi}_i(t) \hat{\phi}_j(t) \rangle$ at different times. In the one-dimensional case, we choose $i = 50$ and plot the correlation as a function of $j$. In the two-dimensional case, on the other hand, we choose $i$ to be the oscillator in the $25^{\text{th}}$ row and $25^{\text{th}}$ column of our $50 \times 50$ lattice and take $j$ to run over all oscillators in the same row. At the initial time, we see in both cases that the self-correlation is the largest, and the correlation function quickly decays beyond nearest neighbors. As time evolves, long-range correlations start to develop in the absence of disorder, as expected. In the one-dimensional case, we find that they grow until $t/\epsilon \approx N/2$ and start to decrease back to their initial value after that, while in the two-dimensional case, they persist even at much later times, similar to the behavior of entanglement entropy shown in fig.\ \ref{fig:2}. In the presence of disorder, we find instead that spatial correlations are suppressed and, in fact, decay faster than in the initial state, again suggesting that the system has localized.

\section{Frequency gap ratio}
\label{sec:fgr}

In this section, we define a measure of localization based on the spectral gap ratio of \cite{Oganesyan:2007} to understand how our results vary with disorder strength and system size. The spectral gap ratio probes correlations between the {\it energy eigenvalues} of the Hamiltonian and allows one to differentiate between the chaotic regime, where adjacent gaps between the eigenvalues are correlated and distributed according to a Gaussian orthogonal ensemble (GOE), and the localized regime, where the gaps are uncorrelated and follow a Poisson distribution. Since the many-body system we consider has an infinite dimensional Hilbert space, we define a frequency gap ratio that probes correlations between the {\it phase space eigenvalues} of the Hamiltonian instead. From an ordered set of symplectic eigenvalues $\{\omega_i\}$ of the Hamiltonian matrix $\boldsymbol{V}$, we first find the gaps between successive eigenvalues $\delta_i = \omega_{i+1}-\omega_i$, and then define the frequency gap ratio as
\bea
    r_{\omega} & = & \overline{{\frac{\min\{\delta_i,\delta_{i+1}\}}{\max\{\delta_i,\delta_{i+1}\}}}} \, ,
\label{eq:gapratio}
\eea
where the bar denotes an average over all gaps and random realizations. We expect this to be a reliable measure of localization for our system since its dynamics are described fully by its frequency spectrum.

\begin{figure*}[!t]
\begin{center}
    \includegraphics[scale=0.55]{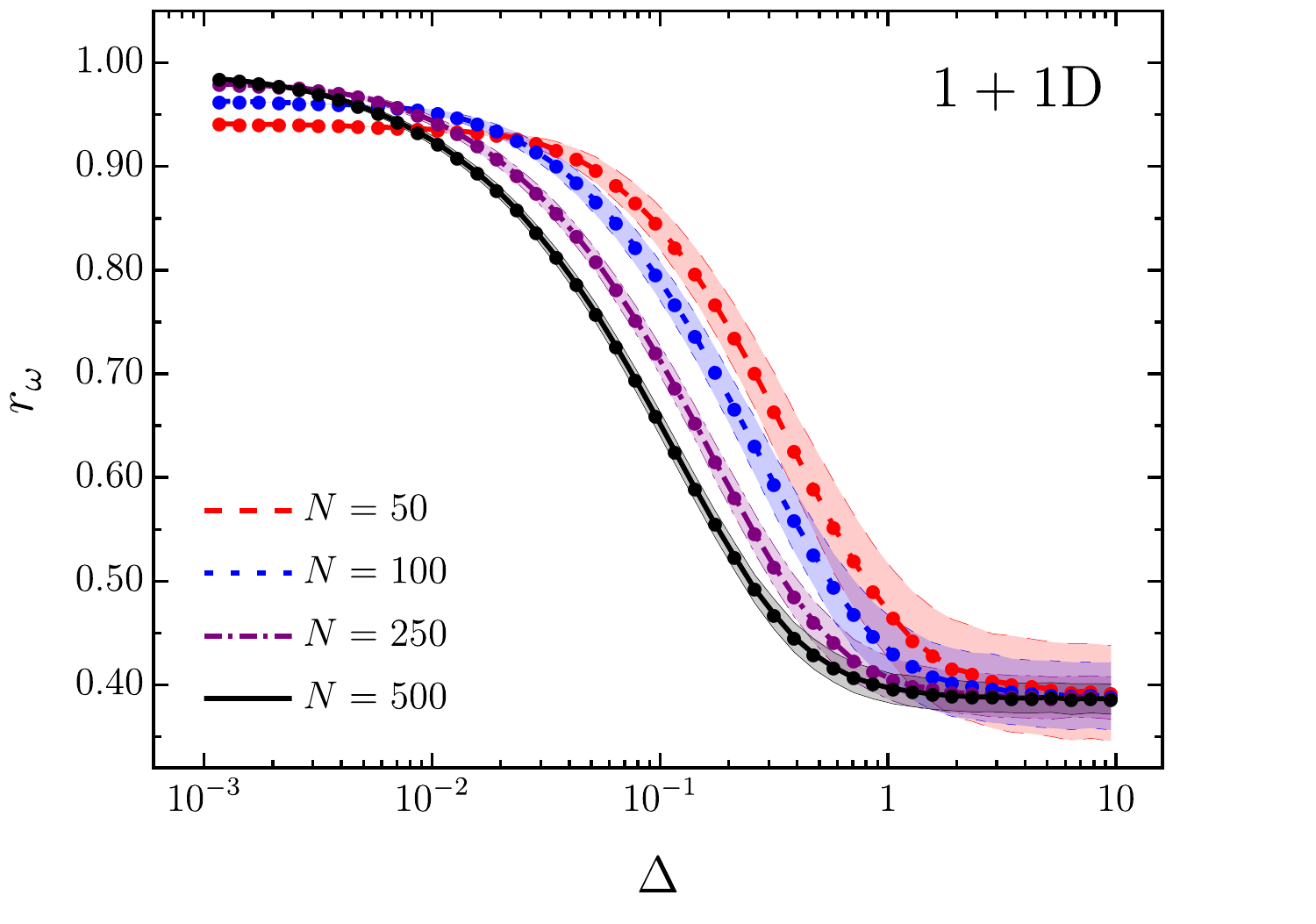}
    \includegraphics[scale=0.55]{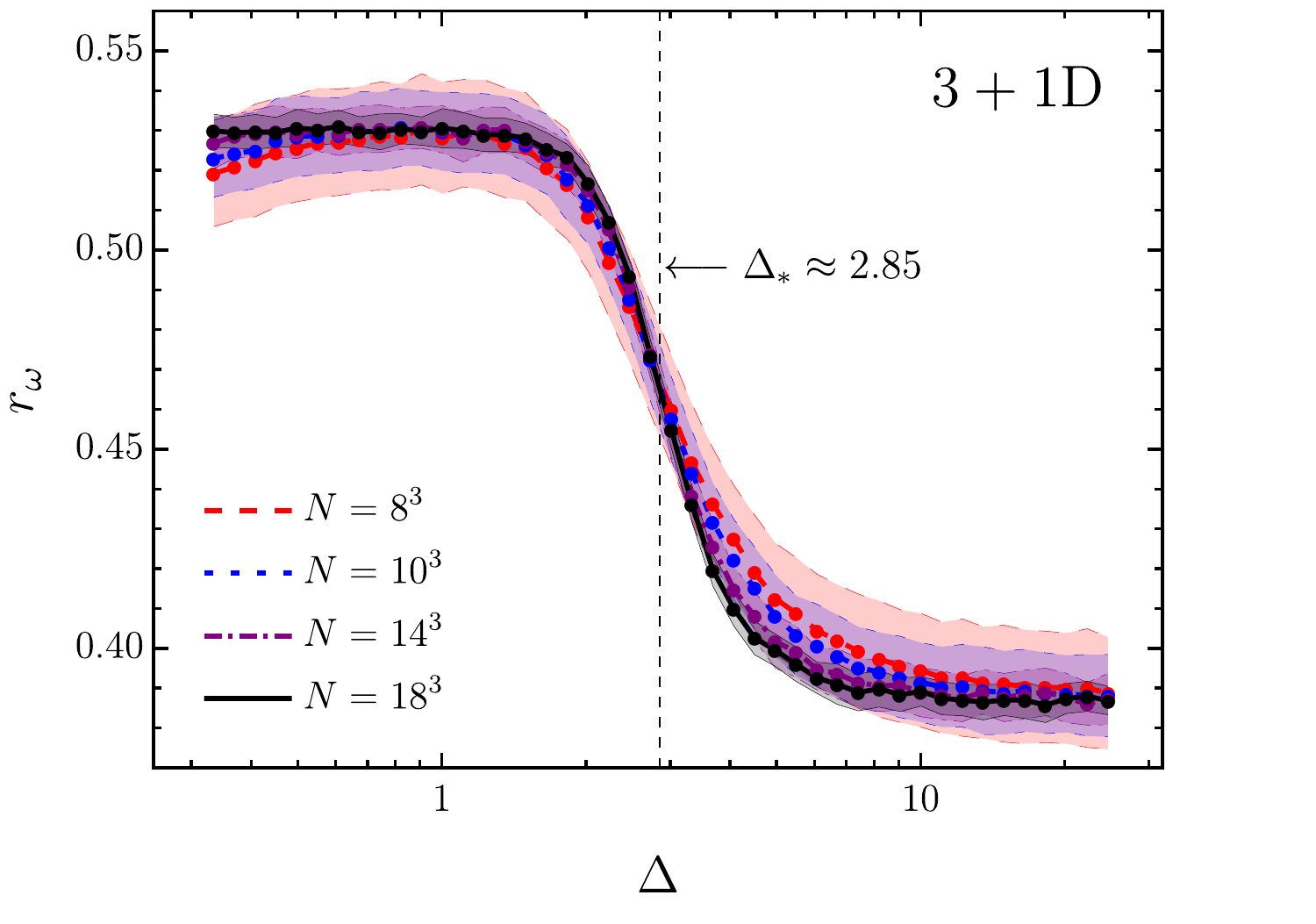} 
    \\
    \includegraphics[scale=0.55]{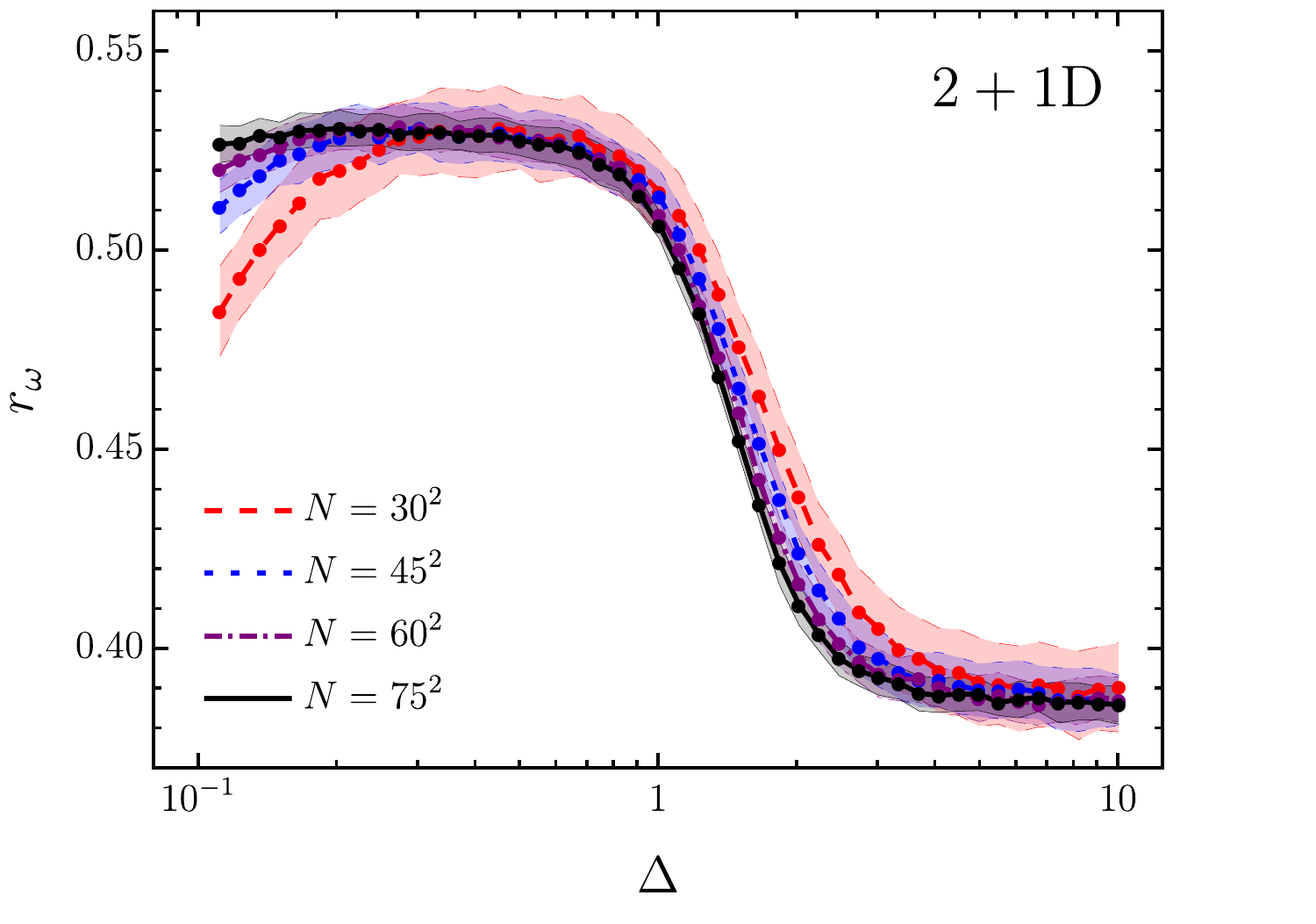}
    \includegraphics[scale=0.55]{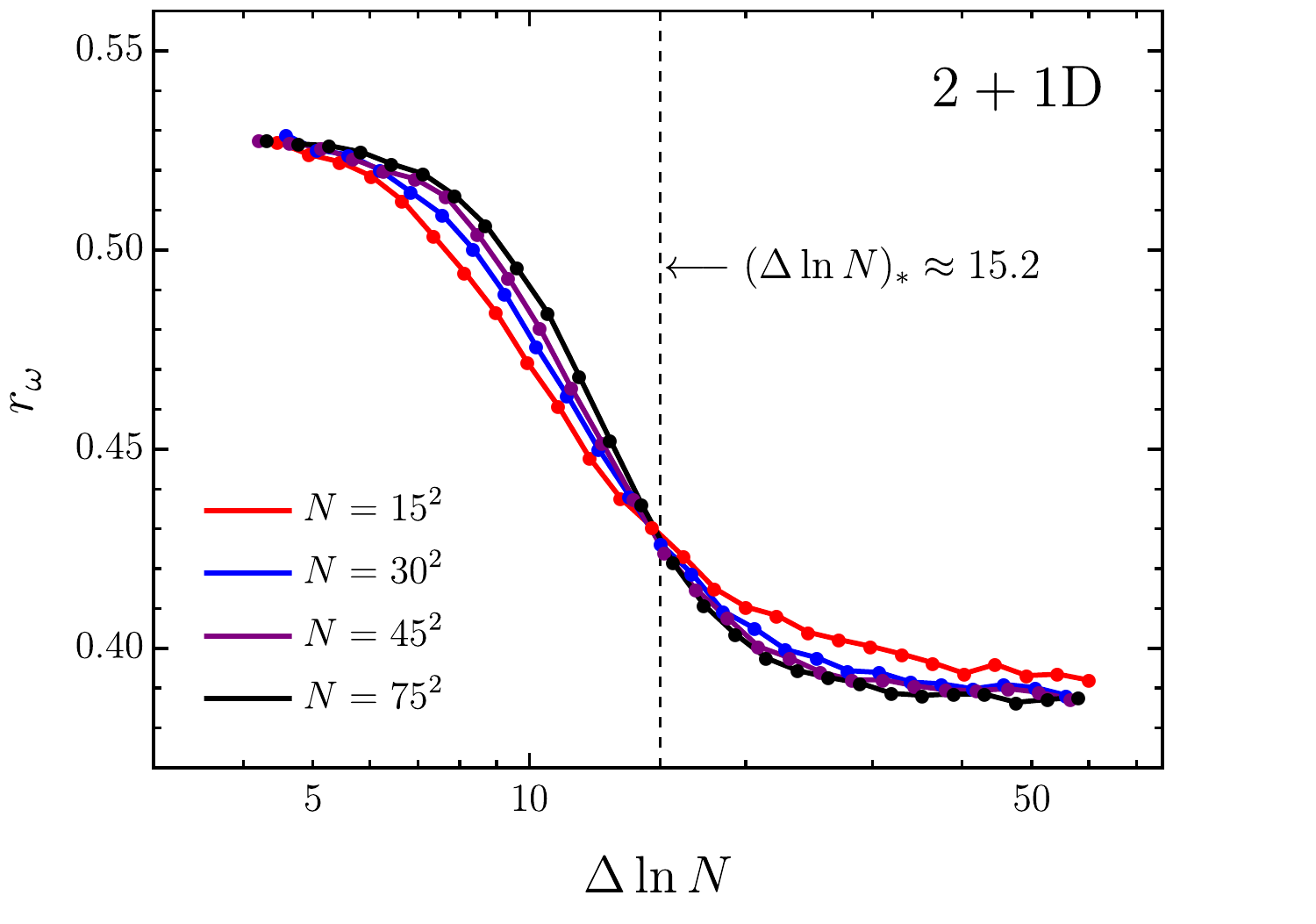}
    \caption{Frequency gap ratio as a function of disorder strength and system size. (Top, left) For a chain of oscillators with $N \in \{50, 100, 250, 500\}$ and each data point being the average over $4000$, $3000$, $2000$, and $1000$ realizations of the $h_i$, respectively. (Top, right) For a three-dimensional lattice of oscillators with $N \in \{8^3, 10^3, 14^3, 18^3\}$ and each data point being the average over $1000$, $500$, $100$, and $50$ realizations, respectively. (Bottom, left) For a two-dimensional lattice of oscillators with $N \in \{30^2, 45^2, 60^2, 75^2\}$ and each data point being the average over $400$, $200$, $75$, and $50$ realizations, respectively. (Bottom, right) Also for a two-dimensional lattice of oscillators, but plotted against the disorder strength rescaled by $\ln N$, with $N \in \{15^2, 30^2, 45^2, 75^2\}$ and each data point being the average over $500$, $400$, $200$, and $50$ realizations, respectively. The shaded region in the first three graphs is the $1\sigma$ error bar.}
\label{fig:4}
\end{center}
\end{figure*}

In fig.\ \ref{fig:4}, we show how the frequency gap ratio $r_{\omega}$ varies with disorder strength $\Delta$ for different system sizes $N$ and in different dimensions, and find results consistent with Anderson localization. In the one-dimensional case (top, left panel), we find that $r_{\omega}$ is close to unity for small disorder and relaxes close to the Poisson value of $2\ln2 -1 \approx 0.39$ as we increase the disorder strength. We also find that the amount of disorder needed for Poisson distributed eigenvalues {\it decreases} as we increase the system size, suggesting that the system would localize in the thermodynamic/continuum limit, where $N \rightarrow \infty$, for arbitrarily small disorder. In the bottom two panels of fig.\ \ref{fig:4}, we show the result in the two-dimensional case. We first find that the gap ratio transitions through a region where $r_{\omega} \approx 0.53$, that coincides with the GOE value, before relaxing to the Poisson value, indicating the presence of a chaotic regime at intermediate disorder strengths. The point at which the transition occurs, however, is known to drift logarithmically to lower disorder strengths as one increases the system size. In the bottom, right panel, we next plot the gap ratio as a function of $\Delta$ scaled by $\ln N$ instead, as discussed in \cite{Suntajs:2022ldk}, for different system sizes, and see that the graphs intersect at the point $(\Delta \ln N)_* \approx 15.2$. This suggests that in the limit of $N \rightarrow \infty$, the transition from GOE to Poisson distributed eigenvalues would occur at $\Delta \rightarrow 0$ and the system would, therefore, again localize for arbitrarily small disorder.

Lastly, we consider the three-dimensional case (top, right panel of fig.\ \ref{fig:4}), where we define the Hamiltonian matrix using eqs.\ (\ref{eq:V1d}) and (\ref{eq:Kmat}) as before, except with $\deg(v_i) = 6$ and an additional pair of nearest-neighbor interactions. We find that the gap ratio transitions from the GOE value to the Poisson one as we increase the disorder strength, similar to the case of two spatial dimensions. Unlike the two-dimensional case, however, the graphs for different system sizes intersect at a critical disorder strength $\Delta_* \approx 2.85$. This suggests that in the continuum limit, there is a real transition between the thermalizing and localizing regime at a finite value of the disorder, unlike the one- and two-dimensional cases, consistent with Anderson localization in $3$D \cite{Suntajs:2021qqr}. We also note that $r_{\omega}$ goes to zero at small disorder strengths in two and three spatial dimensions due to near-degeneracies in $\omega_i$ at small but non-zero $\Delta$, that turn into exact degeneracies and, therefore, vanishing $\delta_i$ in the $\Delta = 0$ case.

\section{Discussion}
\label{sec:disc}

In a scalar QFT that reduces to a many-body system of quantum harmonic oscillators upon discretization, we first showed that subregions thermalize following a global mass quench. This is characterized by a linear growth of entanglement entropy, a transition to an extensive dependence on system size, and the appearance of long-range correlations. We next showed that in the presence of sufficiently large disorder, the entanglement entropy of subregions maintains its initial area law behavior and the system no longer develops long-range correlations in both one and two spatial dimensions. We further defined a frequency gap ratio that measures correlations between adjacent gaps in the phase space eigenvalues of the Hamiltonian. We used it to demonstrate that in the continuum limit for one and two spatial dimensions, arbitrarily small disorder can lead to localization, even though the two-dimensional case exhibits a chaotic regime for small disorder and finite system sizes. For three spatial dimensions, however, the spectral gap ratio suggests a real transition from a thermalizing to localizing regime at a finite value of the disorder. These results are consistent with Anderson localization observed in different systems.

Before we end the paper, we would like to comment on a possible source of the disorder term since the one that we added in eq.\ (\ref{eq:hamdis}) breaks Lorentz invariance in the continuum limit. Consider a scalar QFT on a curved background with a spatially inhomogeneous time component of the metric tensor, i.e., with $g_{00} = -\sigma^2(\vx)$, $g_{0i}= 0$, and $g_{ij} = \delta_{ij}$, where $0$ denotes the time component, $i$ and $j$ denote spatial components, and $\sigma(\vx)$ is some function. The Hamiltonian for this theory would be similar to that written in eq.\ (\ref{eq:ham}), but with an additional factor of $\sigma(\vx)$ multiplying the integrand. The discretized Hamiltonian will then contain a term similar to that in eq.\ (\ref{eq:hamdis}), along with a spatially-inhomogeneous nearest-neighbor coupling term that may lead to localization in the QFT. It may, therefore, be possible for gravity to source the required disorder, at least in $1+1$D and $2+1$D, where infinitesimal disorder is sufficient to localize in the continuum, and it would be interesting to explore this further. It would also be interesting to explore whether there is a connection between the results presented here and the random matrix theory behavior seen in interacting QFTs \cite{Delacretaz:2022ojg}.


\acknowledgments

We especially thank Sarang Gopalakrishnan for comments on a previous version of this paper. We also thank Thomas Hartman, Mark Hertzberg, Archana Kamal, Albion Lawrence, Sarah Shandera, and Brian Swingle for useful conversations. This work was supported in part by the Department of Energy under award number DE-SC0020360.


\appendix
\section{Phase space diagonalization}
\label{app:diag}

In this appendix, we describe in detail the diagonalization method used in the paper to perform exact calculations. It follows a similar treatment as can be found in, for example, \cite{Gurarie:2003,Pirandola:2009}. We consider a quadratic Hamiltonian of $N$ coupled harmonic oscillators of the form 
\bea
    \hat{H} & = & \frac{1}{2\epsilon} \hat{\boldsymbol{\chi}}^{T} \boldsymbol{V} \hat{\boldsymbol{\chi}} \ ,
\label{eq:Hamapp2}
\eea
where $\hat{\boldsymbol{\chi}} = \icol{\hat{\boldsymbol{\phi}} \\ \hat{\boldsymbol{\pi}}}$ is the phase space vector in the physical basis, with
\bea
    \hat{\boldsymbol{\phi}} \, = \, \[
    \begin{array}{c}
		\hat{\phi}_1 \\
		\vdots \\
		\hat{\phi}_N
	\end{array} \] \ \ {\rm and} \quad
    \hat{\boldsymbol{\pi}} \, = \, \[
    \begin{array}{c}
		\hat{\pi}_1 \\
		\vdots \\
		\hat{\pi}_N
	\end{array} \] ,
\eea
and $\boldsymbol{V}$ is a $2N \times 2N$ symmetric matrix whose diagonal terms constitute the free part of the Hamiltonian and off-diagonal terms the couplings between different oscillators and momenta. We include a factor of $\epsilon$ in the Hamiltonian to keep $\hat{\phi}_i$ and $\hat{\pi}_i$ dimensionless, consistent with eqs.\ (\ref{eq:Hamapp}) and (\ref{eq:BoldPhiPi}), but note that the diagonalization procedure can easily be adapted to harmonic oscillators with standard dimensions. In this construction, the commutation relations can be written as
\bea
    [\hat{\chi}_{A},\hat{\chi}_{B}] & = & iJ_{AB} \ ,
\label{eq:comm} \\
    J_{AB} & = & \begin{bmatrix} 0 & \boldsymbol{I}_{N} \\ -\boldsymbol{I}_{N} & 0 \end{bmatrix}_{AB} \, ,
\label{eq:SSM}
\eea
where the indices $A$ and $B$ run from $1$ to $2N$ and $\boldsymbol{I}_{N}$ is the $N$-dimensional identity matrix.\footnote{We note that a slightly different structure is also used in the literature, where $\hat{\boldsymbol{\chi}} = \big(\hat{\phi}_1, \hat{\pi}_1, \dots, \hat{\phi}_N, \hat{\pi}_N\big)$ and the commutator becomes $[\hat{\chi}_{A}, \hat{\chi}_{B}] =  -\left( \boldsymbol{I}_N \otimes \boldsymbol{\sigma}_y \right)_{AB}$, instead of that in eqs.\ (\ref{eq:comm}) and (\ref{eq:SSM}), which can be rewritten as $[\hat{\chi}_{A}, \hat{\chi}_{B}] = - \left( \boldsymbol{\sigma}_y \otimes \boldsymbol{I}_N \right)_{AB}$.}

We assume that $\boldsymbol{V}$ is a positive matrix so that the Hamiltonian is bounded from below, but otherwise allow for arbitrary couplings between any oscillators and/or conjugate momenta. This allows us to use the diagonalization procedure developed by Williamson \cite{Williamson:1936} -- for any $2N\times2N$ symmetric positive-definite matrix $\boldsymbol{V}$, there exists a diagonal matrix $\boldsymbol{W}$ and symplectic matrix $\boldsymbol{M}$ such that
\bea
    \boldsymbol{W} & = & (\boldsymbol{M}^{-1})^{T} \boldsymbol{V} \boldsymbol{M}^{-1} \ = \ \begin{bmatrix} \boldsymbol{\Omega} & 0 \\ 0 & \boldsymbol{\Omega} \end{bmatrix} ,
\label{eq:W} \\
    \boldsymbol{\Omega} & = & \textnormal{diag}(\{\omega_i\})\, ,
\label{eq:freqmat}
\eea
where $\{\omega_i\}$ is the set of $N$ symplectic eigenvalues of $\boldsymbol{V}$. Using this, we can rewrite the Hamiltonian in the decoupled form
\bea
    \hat{H} & = & \frac{1}{2\epsilon} \hat{\boldsymbol{\chi}}_{D}^{T} \boldsymbol{W} \hat{\boldsymbol{\chi}}_D \nn \\
    & = & \frac{1}{2\epsilon} \sum_{i=1}^N \omega_i \big( \hat{\pi}_{D,i}^2 + \hat{\phi}_{D,i}^2 \big) \, ,
\label{eq:Hamdiag} 
\eea
where $\hat{\boldsymbol{\chi}}_{D} = \boldsymbol{M} \hat{\boldsymbol{\chi}}$ is the phase space vector in the decoupled basis. For this transformation to preserve the commutator in eq.\ (\ref{eq:comm}), it has to obey
\bea
    \boldsymbol{M}^{T} \boldsymbol{J} \boldsymbol{M} & = & \boldsymbol{J} \, , 
\label{eq:sympl}
\eea
which is guaranteed since $\boldsymbol{M}$ is a symplectic matrix. 

We now describe the procedure to find the transformation matrix $\boldsymbol{M}$. The steps below can be followed to find the Williamson normal form and matrix $\boldsymbol{M}$ for any quadratic Hamiltonian.
\begin{enumerate}
    \item Construct the Williamson form $\boldsymbol{W}$ using the symplectic eigenvalues of $\boldsymbol{V}$. The set of symplectic eigenvalues can be obtained by calculating the eigenvalues of $i\boldsymbol{J} \boldsymbol{V}$ and selecting the positive ones.
\label{diagstep1}
    \item Find $\boldsymbol{V}^{1/2}$ using orthogonal diagonalization. $\boldsymbol{V}^{1/2}$ is a real, symmetric matrix, such that $\boldsymbol{V}^{1/2}\boldsymbol{V}^{1/2} = \boldsymbol{V}$.
\label{diagstep2}
    \item Define $\boldsymbol{M} = \boldsymbol{W}^{-1/2} \boldsymbol{R} \boldsymbol{V}^{1/2}$, where $\boldsymbol{R}$ is an orthogonal matrix, so that eq.\ (\ref{eq:W}) is satisfied.
\label{diagstep3}
    \item Find the orthogonal matrix $\boldsymbol{R}$ such that $\boldsymbol{M}$ obeys eq.\ (\ref{eq:sympl}). We rewrite that condition as $\boldsymbol{Y} = \boldsymbol{R} \boldsymbol{X} \boldsymbol{R}^{T}$, where we define $\boldsymbol{X} = \boldsymbol{V}^{-1/2} \boldsymbol{J} \boldsymbol{V}^{-1/2}$ and $\boldsymbol{Y} = \boldsymbol{W}^{-1/2} \boldsymbol{J} \boldsymbol{W}^{-1/2}$. 
\label{diagstep4}
\end{enumerate}
We clarify that in step \ref{diagstep1}, the matrix $i \boldsymbol{J} \boldsymbol{V}$ has eigenvalues $\{\omega_i\}$ and $\{-\omega_i\}$. This is the simplest way to find the frequencies needed to define $\boldsymbol{W}$ and how we obtain the symplectic eigenvalues of submatrices of $\boldsymbol{\Gamma}$ needed for the entanglement entropy calculations in the main text. For step \ref{diagstep4}, $\boldsymbol{R}$ can be obtained by first finding the unitary matrix $\boldsymbol{U}$ which diagonalizes $\boldsymbol{X}$, such that
\bea
    \boldsymbol{U} \boldsymbol{X} \boldsymbol{U}^{\dagger} & = & -i \bigoplus_{i=1}^N \begin{bmatrix} x_i & 0 \\ 0 & -x_i \end{bmatrix} ,
\eea
where $x_i$ are the eigenvalues. We can then directly define
\bea
    \boldsymbol{R} & = & \boldsymbol{\Sigma} \boldsymbol{U} ,
\eea
with the matrix $\boldsymbol{\Sigma}$ having the form
\bea
    \boldsymbol{\Sigma} & = & \sum_{i = 1}^{N} \Big[ i \boldsymbol{x}_i \otimes \boldsymbol{\psi}_i^{-} + \boldsymbol{x}_{N+i} \otimes \boldsymbol{\psi}_i^{+} \Big] \, ,
\eea
where $\boldsymbol{x}_i$ is the $i^{\rm th}$ cartesian vector on $\mathbb{R}^{2N}$ and $\boldsymbol{\psi}_{i}^{\pm}$ is a $2N$-dimensional vector that is zero except for the $(2i-1)^{\rm th}$ component that is $1$ and the $(2i)^{\rm th}$ component that is $\pm 1$. The matrix $\boldsymbol{\Sigma}$ is structure-dependent, and it brings the eigenvalues of $\boldsymbol{X}$ to the two block off-diagonals in order to match with $\boldsymbol{Y}$ as needed for step \ref{diagstep4}.


\section{Time evolution}
\label{app:td}

In this appendix, we describe the procedure to calculate the exact time evolution of the system once we have found the diagonalization matrix $\boldsymbol{M}$ using the procedure described in appendix\ \ref{app:diag}. Since we restrict to Gaussian initial states and quadratic Hamiltonians, we can focus on finding the time evolution of the covariance matrix $\boldsymbol{\Gamma}$, that we use to calculate the entanglement entropy. We start by restating the definition of $\boldsymbol{\Gamma}$ given in eq.\ (\ref{eq:gammadef}),
\bea
    \Gamma_{AB}(t) & = & \frac{1}{2} \big\langle \left\{ \hat{\chi}_A(t), \hat{\chi}_B(t) \right\} \big\rangle -\big\langle \hat{\chi}_A(t)\big\rangle\big\langle \hat{\chi}_B(t) \big\rangle \, , \nn\\
\label{eq:Gammadef} 
\eea
where $\{\cdot,\cdot\}$ is the anti-commutator. Let us denote the covariance matrix at the initial time and in the original operator basis of the Hamiltonian in eq.\ (\ref{eq:Hamapp2}), which we refer to as the physical basis, with $\boldsymbol{\Gamma}(0)$. Using eq.\ (\ref{eq:Gammadef}) along with the transformation matrix $\boldsymbol{M}$, we can find the initial covariance matrix in the decoupled basis,
\bea
    \boldsymbol{\Gamma}_{D}(0) & = & \boldsymbol{M} \boldsymbol{\Gamma}(0) \boldsymbol{M}^{T} .
\label{eq:Gammadb}
\eea
Note that the subscript in $\boldsymbol{\Gamma}_{D}(0)$ denotes the covariance matrix in the decoupled basis of the Hamiltonian, not that it is itself diagonal.

The time evolution of $\hat{\boldsymbol{\chi}}_D(t)$ is described by the Heisenberg equation of motion,
\bea
    \frac{\d \hat{\boldsymbol{\chi}}_D}{\d t} & = & i\big[ \hat{H}, \hat{\boldsymbol{\chi}}_D \big] \ = \ \frac{1}{\epsilon} \boldsymbol{J} \boldsymbol{W} \hat{\boldsymbol{\chi}}_D \, ,
\eea
where we have used the Hamiltonian in the form of eq. (\ref{eq:Hamdiag}) and the commutation relations from eq.\ (\ref{eq:comm}) in the second equality. The solution to this equation is simply
\bea
    \hat{\boldsymbol{\chi}}_D(\tau) & = & e^{\boldsymbol{J} \boldsymbol{W} \tau} \hat{\boldsymbol{\chi}}_D(0) \, ,
\label{eq:chitd}
\eea
where we have introduced a dimensionless time parameter $\tau = t/\epsilon$. Note that since $[\boldsymbol{J}\boldsymbol{W},\boldsymbol{J}] = 0$ and $\big( e^{\boldsymbol{J} \boldsymbol{W} \tau} \big)^T = e^{-\boldsymbol{J} \boldsymbol{W} \tau}$, the time evolution matrix also satisfies eq.\ (\ref{eq:sympl}) and is therefore symplectic. From eq.\ (\ref{eq:chitd}) and the definition of the covariance matrix in eq.\ (\ref{eq:Gammadef}), it follows that
\bea
    \boldsymbol{\Gamma}_D(t) & = & e^{\boldsymbol{J} \boldsymbol{W} \tau} \boldsymbol{\Gamma}_{D}(0) e^{-\boldsymbol{J} \boldsymbol{W} \tau} \, .
\label{eq:Gammatd}
\eea
Finally, we can transform back to find the covariance matrix as a function of time in the physical basis,
\bea
    \boldsymbol{\Gamma}(\tau) & = & \boldsymbol{M}^{-1} \boldsymbol{\Gamma}_D(\tau) (\boldsymbol{M}^{-1})^{T} \, .
\label{eq:Gammapbtd}
\eea
In the main text, we use submatrices of the resulting $\boldsymbol{\Gamma}(\tau)$ to find how the entanglement entropy of subregions and correlation functions evolve in time.


\bibliographystyle{apsrev}
\bibliography{references}


\end{document}